\documentclass[twocolumn,showpacs,superscriptaddress,amsmath,amssymb]{revtex4}
\usepackage{graphicx}
\usepackage{dcolumn}
\usepackage{bm}
\usepackage{color}
\usepackage{mathrsfs}
\usepackage{ulem}

\def\bd{
\begin{document}} \def\ed{\end{document}}
\def\bmp{\begin{minipage}} \def\emp{\end{minipage}}
\def\bcc{\begin{center}} \def\ecc{\end{center}}     \def\npg{\newpage}
\def\beq{\begin{equation}} \def\eeq{\end{equation}} \def\hph{\hphantom}
\def\be{\begin{equation}} \def\ee{\end{equation}} \def\r#1{$^{[#1]}$}
\def\n{\noindent} \def\ni{\noindent} \def\pa{\parindent}
\def\hs{\hskip} \def\vs{\vskip} \def\hf{\hfill} \def\ej{\vfill\eject}
\def\cl{\centerline} \def\ob{\obeylines}  \def\ls{\leftskip}
\def\underbar#1{$\setbox0=\hbox{#1} \dp0=1.5pt \mathsurround=0pt
   \underline{\box0}$}   \def\ub{\underbar}    \def\ul{\underline}
\def\f{\left} \def\g{\right} \def\e{{\rm e}} \def\o{\over} \def\d{{\rm d}}
\def\vf{\varphi} \def\pl{\partial} \def\cov{{\rm cov}} \def\ch{{\rm ch}}
\def\la{\langle} \def\ra{\rangle} \def\EE{e$^+$e$^-$} \def\pt{p_{\rm t}}
\def\pti{p_{{\rm t},i}} \def\vti{v_{{\rm t},i}}
\def\ptj{p_{{\rm t},j}}\def\Pt{P_{\rm t}} \def\vt{v_{\rm t}}

\def\bitz{\begin{itemize}} \def\eitz{\end{itemize}}
\def\btbl{\begin{tabular}} \def\etbl{\end{tabular}}
\def\btbb{\begin{tabbing}} \def\etbb{\end{tabbing}}
\def\beqar{\begin{eqnarray}} \def\eeqar{\end{eqnarray}}
\def\\{\hfill\break} \def\dit{\item{-}} \def\i{\item}
\def\bbb{\begin{thebibliography}{9} \itemsep=-1mm}
\def\ebb{\end{thebibliography}} \def\bb{\bibitem}
\def\bpic{\begin{picture}(260,240)} \def\epic{\end{picture}}
\def\akgt{\cl{\bf ACKNOWLEDGMENTS}}
\def\fgn{\noindent{\bf\large\bf figure captions}}
\def\m1{\langle N_p\rangle} \def\u2{\langle N_{\bar p}\rangle} \def\Nap{N_{\bar
p}}
\def\lan{\langle}
\def\ran{\rangle}
\def\p{\pi}
\def\ifmath#1{\relax\ifmmode #1\else $#1$\fi}%
\def\rc{\ifmath{{\mathrm{c}}}}
\def\cut{\ifmath{{\mathrm{cut}}}}
\def\rF{\ifmath{{\mathrm{F}}}}
\def\rK{\ifmath{{\mathrm{K}}}}
\def\rp{\ifmath{{\mathrm{p}}}}
\def\rt{\ifmath{{\mathrm{t}}}}
\def\LAB{\ifmath{{\mathrm{LAB}}}}
\def\cut{\ifmath{{\mathrm{cut}}}}
\def\beq{\begin{equation}}
\def\eeq{\end{equation}}

\newcommand{\cinst}[2]{$^{\mathrm{#1}}$~#2\par}
\newcommand{\crefi}[1]{$^{\mathrm{#1}}$}
\newcommand{\crefii}[2]{$^{\mathrm{#1,#2}}$}
\newcommand{\crefiii}[3]{$^{\mathrm{#1,#2,#3}}$}
\newcommand{\HRule}{\rule{0.5\linewidth}{0.5mm}}

\bd
\title{Finite-size behaviour of generalized susceptibilities in the whole phase plane of the Potts model}

\author{Xue Pan} 
\affiliation{School of Electronic Engineering, Chengdu Technological University, Chengdu 611730, China}
\author{Yanhua Zhang}
\affiliation{Key Laboratory of Quark and Lepton Physics (MOE) and Institute of Particle Physics, Central China Normal University, Wuhan 430079, China }
\author{Lizhu Chen} 
\affiliation{ School of Physics and Optoelectronic Engineering, Nanjing University of Information Science and Technology, Nanjing 210044, China}
\author{Mingmei Xu} 
\affiliation{Key Laboratory of Quark and Lepton Physics (MOE) and Institute of Particle Physics, Central China Normal University, Wuhan 430079, China  }
\author{Yuanfang Wu} 
\affiliation{Key Laboratory of Quark and Lepton Physics (MOE) and Institute of Particle Physics, Central China Normal University, Wuhan 430079, China }

\begin{abstract}
We study the sign distribution of generalized magnetic susceptibilities in the temperature-external magnetic field plane using the three-dimensional three-state Potts model. We find that the sign of odd-order susceptibility is opposite in the symmetric (disorder) and broken (order) phases, but that of the even-order one remains positive when it is far away from the phase boundary. When the critical point is approached from the crossover side,  negative fourth-order magnetic susceptibility is observable. It is also demonstrated that  non-monotonic behavior occurs in the temperature dependence of the generalized susceptibilities of the energy. The finite-size scaling behavior of the specific heat in this model is mainly controlled by the critical exponent of the magnetic susceptibility in the three-dimensional Ising universality class.

\end{abstract}


\maketitle

\section{Introduction}

One of the main goals of heavy ion experiments is to locate the critical point, and/or the boundary of the quantum chromodynamics (QCD) phase transition~\cite{main goal}. An expected QCD phase diagram in the temperature and baryon chemical potential plane is sketched in Fig.~1(a), where the system undergoes a first-order phase transition at high baryon chemical potential ($\mu_B$) and low temperature~\cite{confinement first 1, confinement first 2, confinement first 3}. With the decrease of $\mu_B$ and increase of temperature, the first-order phase transition line is expected to end at a critical point in some theories, in the three-dimensional Ising universality class~\cite{Stephanov-PRL81, PRL105}, although no critical point has yet been found in 2+1 lattice QCD~\cite{PRD95-karsch}. At high temperature and vanishing chemical potential, the calculations of lattice QCD have shown it is a crossover~\cite{nature-crossover}.

To probe this expected phase diagram in heavy ion experiments, the high-order cumulants of the conserved charges are suggested as sensitive observables of the critical fluctuations~\cite{Stephanov-PRL81,stephnov-prd1999,koch,MCheng-PRD79,STAR-PRL112,STAR-PRL113}. Theoretically, they correspond to the generalized susceptibilities of the conserved charges, and can be calculated by lattice QCD at vanishing $\mu_B$~\cite{MCheng-PRD79,Karsch-CEJP10,Karsch-PRL109, Fodor-PRD92,Fodor-PRL111}, and QCD effective models at finite $\mu_B$~\cite{Asakawa-prl103,WJF-PRD82}. Peak-like structures or oscillations occur in the temperature dependence of high-order susceptibilities. So non-monotonic behavior of high-order cumulants of conserved charges is considered as a signal of the critical point~\cite{Stephanov-PRL107,Karsch-EPJC71,DengJ-PRD9395,Xue-NPA913}.

If the QCD critical point exists, it should be in the three-dimensional Ising universality class. Based on the universality of the critical behavior, it has been pointed out that the fourth-order cumulant of the order parameter is negative when the critical point is approached on the crossover side in this universality class ~\cite{Stephanov-PRL107}.

The three-dimensional three-state Potts model shares the global Z(3) symmetry with the finite temperature QCD with infinitely heavy quarks~\cite{Z3-NPB243}. The temperature driven phase transition to a broken Z(3) symmetry phase corresponds to the first order deconfining phase transition in QCD~\cite{first order 1,first order 2}. When an external magnetic field is turned on in the Potts model, the Z(3) symmetry is explicitly broken. As the magnetic field increases, the first order phase transition is weakened and ends at a critical point $(\beta_{c}, h_{c})$ = 0.54938(2), 0.000775(10)), which is in the three-dimensional Ising universality class~\cite{Z3-NPB243,Karsch-PLB488}. Beyond the critical point, it is a crossover. It is interesting to study the behavior of the generalized susceptibilities of the magnetization, which is related to the order parameter in this model~\cite{Xue-JPG42}.

The phase diagram of this model in the phase plane of temperature and external field is presented in Fig.~1(b). The first order phase transition line is showed by a solid red line. It separates the whole phase plane into broken (ordered) and symmetric (disordered) phases. The red star represents the critical point. On this phase diagram, the sign distribution of the high-order cumulants is studied. Near the phase boundary, we choose four lines, such as the violet, blue, green and black dashed lines in Fig.~1(b), to analyze the behavior of the generalized susceptibilities of the order parameter when the critical point is approached on the crossover side.

In the vicinity of the critical point, the specific heat, which  corresponds to the second-order susceptibility of the energy, should be divergent in the thermodynamical limit. In this paper, the finite-size behavior of the second to fourth-order susceptibilities of the energy is also studied in the vicinity of the critical point in the three-dimensional three-state Potts model, as the red arrow in Fig.~1(b) shows.

\begin{figure}
	\includegraphics[width=0.23\textwidth]{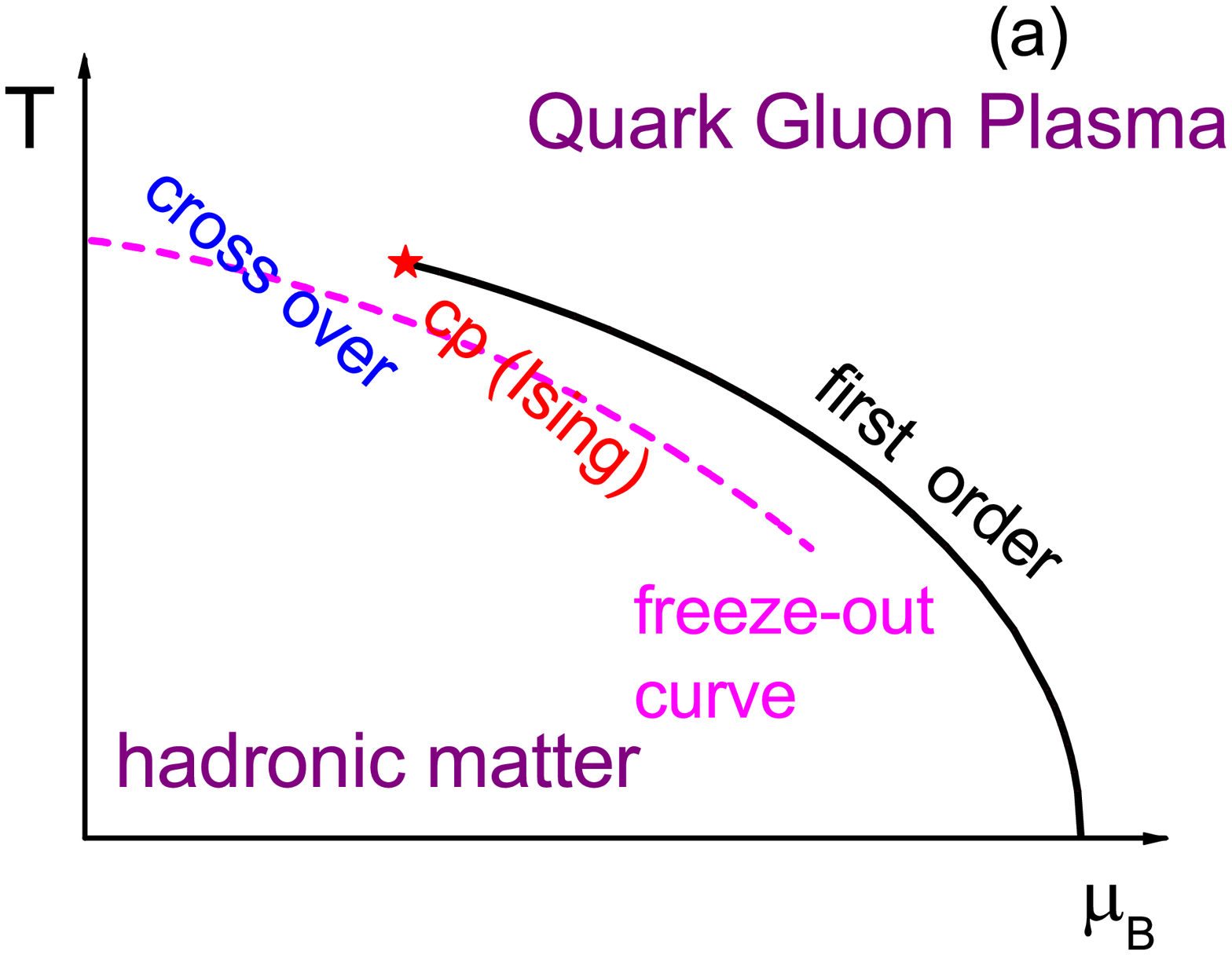}
	\includegraphics[width=0.23\textwidth]{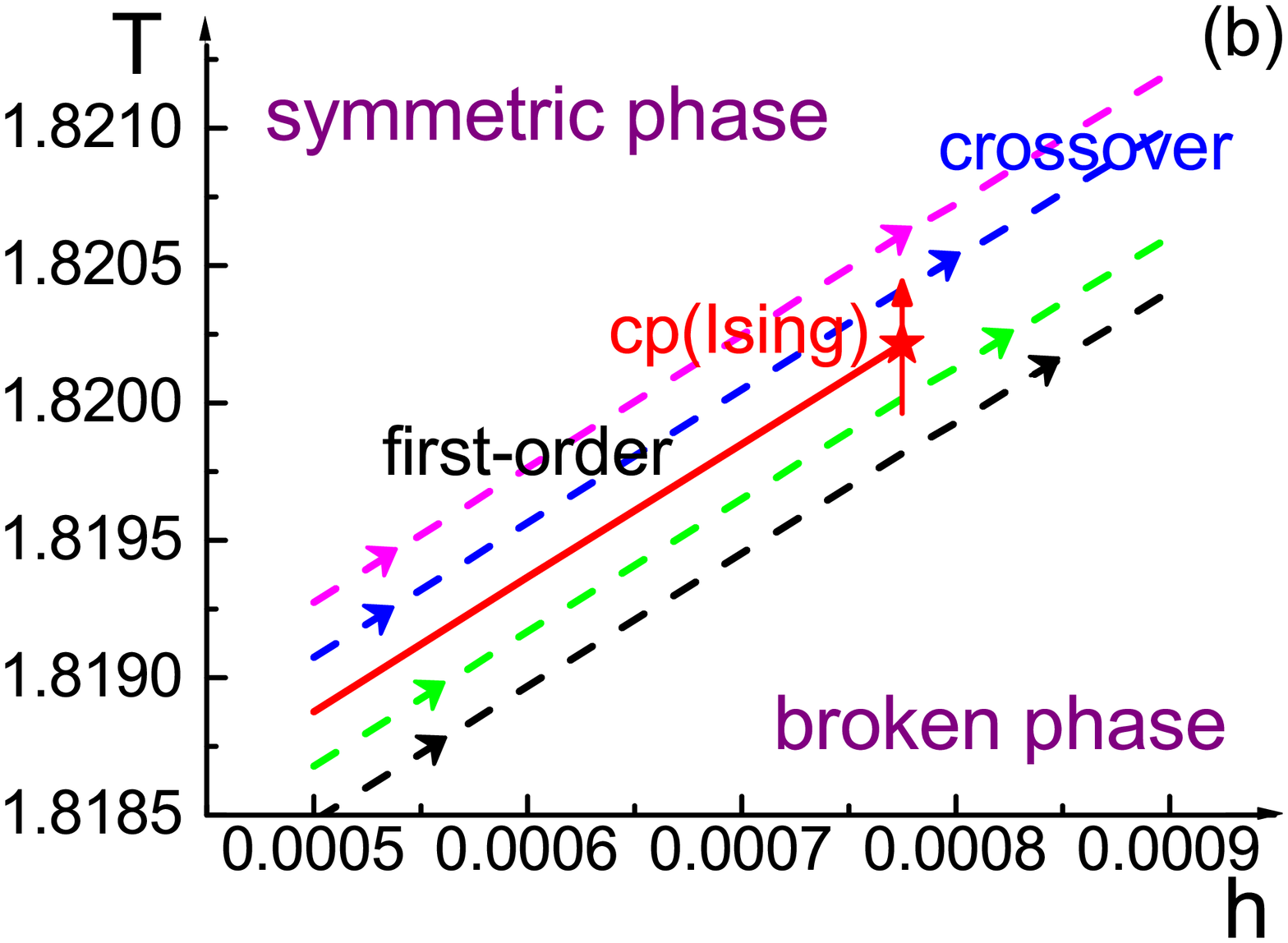}
	\caption{\label{Fig. 1}(color online) Cartoon of phase diagrams of QCD on the temperature-baryon chemical potential plane (a), and the three-dimensional three-state Potts model on the temperature-external magnetic field plane (b).}
\end{figure}

This paper is organized as follows. In Section 2, the formalism of the generalized susceptibilities of the magnetization and energy in the framework of the Potts model is described and derived. In Section 3, the sign distribution of the high-order magnetic susceptibilities is presented and discussed in the temperature-external magnetic field plane. The behavior of the second- to fourth-order magnetic susceptibilities is analyzed when the critical point is approached from the crossover side. The temperature dependence of the second- to fourth-order susceptibilities of the energy in finite-size systems is presented and discussed  in Section 4. Finally, the summary and conclusions are given in Section 5.

\section{Generalized susceptibilities in the framework of the Potts model}

The partition function of the Potts model is defined as~\cite{Z3-NPB243},
\begin{equation}\label{partition func}
Z(\beta, h) =\sum_{\{{s}_{i}\}}e^{-(\beta E-hM)},
\end{equation}
where $s_i \in\{1,2,3\}$ is the spin at site $i$ of a three-dimensional cubic lattice. $\beta=1/T$ is the reciprocal of temperature, and $h=\beta H$ is the normalized external magnetic field. $E$ and $M$ denote the energy and magnetization respectively, i.e.,
\begin{equation}\label{energy}
E=-J\sum_{\langle i,j\rangle}\delta(s_i,s_j), M=\sum_{i}\delta(s_i,s_g).
\end{equation}
$J$ is an interaction energy between nearest-neighbour spins $\langle i,j\rangle$, and is set up as 1 in our
calculations. $s_g$ is the direction of the ghost spin, which the magnetization of non-vanishing external field $h>0$ prefers. The order parameter of the system is defined as~\cite{FY Wu}
\begin{equation}\label{order parameter}
m = \frac{3}{2}\frac{\langle M\rangle}{V}-\frac{1}{2},
\end{equation}
where $V=L^3$ and $L$ is the number of lattice points in each direction. $\left\langle M\right\rangle$ is the mean of all the samples generated.

At vanishing external magnetic field $h=0$, the model is expected to undergo a temperature driven first-order phase transition due to spontaneous symmetry breaking. It is similar to the pure gauge QCD theory, where the corresponding order parameter is the Polyakov loop. It plays the role of magnetization in the spin models, signaling the spontaneous breaking of center symmetry of the deconfined phase.

In the vicinity of the critical point of the three-dimensional three-state Potts model, the original operators for the energy and magnetization, $E$ and $M$, lose their meanings as operators, being conjugate to the temperature-like and symmetry breaking couplings. There is a linear relation of the couplings ($\beta$, $h$) to the temperature and symmetry-breaking couplings ($\tau$, $\zeta$) in the Ising model as follows,
\begin{equation}\label{linear relation}
\tau = \beta-\beta_c + a(h-h_c),~~\zeta = h-h_c + b(\beta-\beta_c).
\end{equation}
$a$ and $b$ are two known mixed parameters~\cite{Karsch-PLB488}.

The susceptibility $\chi_2$ constructed from the magnetization $M$ can be obtained from the second-order derivative of the (reduced) free energy density ($f=-\frac{1}{V}\ln Z$) with respect to the external magnetic field,
\begin{equation}\label{susceptibility}
\left. \chi_2=-\frac{\partial^2 f}{\partial h^2}\right |_T=\frac{1}{V}(\langle M^2\rangle-\langle M\rangle^2).
\end{equation}

Without $1/V$, the right-hand part of Eq.~\eqref{susceptibility} is the second-order cumulant of the magnetization. The high-order derivatives of the free energy density with respect to $h$ are the corresponding generalized magnetic susceptibilities. The $n$th-order magnetic susceptibility is as follows,
\begin{equation}\label{nth order parameter}
\left. \chi_n=-\frac{\partial^n f}{\partial h^n}\right |_T .
\end{equation}
For $n=3$, 4, 5, and 6, we have,
\begin{equation}\label{3 cumulants}
\chi_{3}=\frac{1}{V}{\langle \delta {M}^3 \rangle},
\end{equation}
\begin{equation}\label{4 cumulants}
\chi_{4}=\frac{1}{V}(\langle \delta {M}^4 \rangle-3\langle\delta {M}^2 \rangle^2),
\end{equation}
\begin{equation}\label{5 cumulants}
\chi_{5}=\frac{1}{V}(\langle \delta {M}^5 \rangle-10\langle\delta {M}^3 \rangle \langle\delta {M}^2 \rangle),
\end{equation}
\begin{equation}\label{6 cumulants}
\begin{split}
\chi_{6}=\frac{1}{V}(\langle \delta {M}^6 \rangle-10\langle \delta {M}^3 \rangle^2+30\langle \delta {M}^2 \rangle^3\\-15\langle \delta {M}^4\rangle\langle \delta {M}^2\rangle),
\end{split}
\end{equation}
where $\delta {M}=M-\langle M\rangle$.

Similarly, the generalized susceptibilities of the energy ($\chi_n^E$) can be found from the derivatives of the (reduced) free energy density with respect to $\beta$. The form of $\chi_n^E$ is the same as the $n$th-order susceptibility of the magnetization. Replacing $M$ by $E$ in Eqs.~\eqref{susceptibility}, \eqref{3 cumulants} and \eqref{4 cumulants}, one can get the expressions for $\chi_2^{E}$, $\chi_3^{E}$ and $\chi_4^{E}$, respectively.

In this paper, the estimation of the generalized susceptibilities is based on Monte Carlo simulations of the three-dimensional three-state Potts model,  performed on system sizes of $L^3$ by the Wolff cluster algorithm~\cite{wolff}. The helical boundary conditions were used. For each pair of couplings $(\beta, h)$, we produced a total of 50000 independent configurations.

\section{Fluctuations of generalized magnetic susceptibilities}

The sign distributions of $\chi_{3,4,5,6}$ on the whole $T-h$ plane for system size $L=60$ are presented in Fig.~2(a), 2(b), 2(c), and 2(d), respectively. The green and yellow areas correspond to the positive and negative values of $\chi_{3,4,5,6}$, respectively. The solid red line presents the first-order phase transition line. It ends at the critical point, the red point. The red dashed line indicates the crossover.

\begin{figure}
	\includegraphics[width=0.48\textwidth]{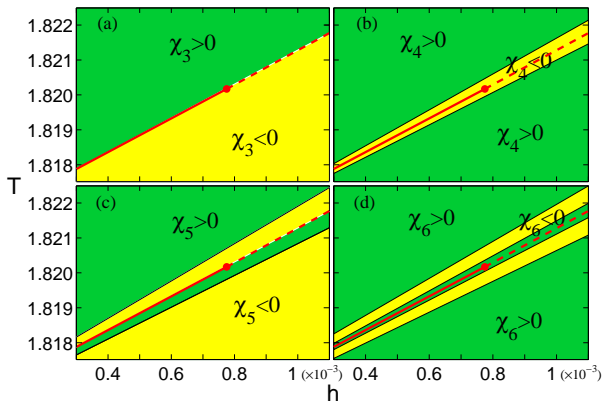}
	\caption{\label{Fig. 2}(color online) The sign distribution of $\chi_{3, 4,5,6}$ on the $T-h$ plane,
		where the green and yellow areas are positive and negative, respectively. The red solid line denotes the phase boundary. }
\end{figure}

In Fig.~2(a), the sign distribution of $\chi_3$ is separated into two parts by the phase boundary. It is positive (green) above the boundary, and negative (yellow) below. Similarly, in Fig.~2(c), the sign distribution of $\chi_5$ is separated into two parts by two narrow bands. One is negative (yellow) above the transition line, and the other is positive (green) below the transition line, so the sign of $\chi_5$ changes three times. One change happens at the phase boundary and the other two are at the lines near the two sides of the boundary. Far away from the phase boundary, the signs of both $\chi_3$ and $\chi_5$ are opposite in symmetry and broken phases.

\begin{figure*}
	\includegraphics[width=0.95\textwidth]{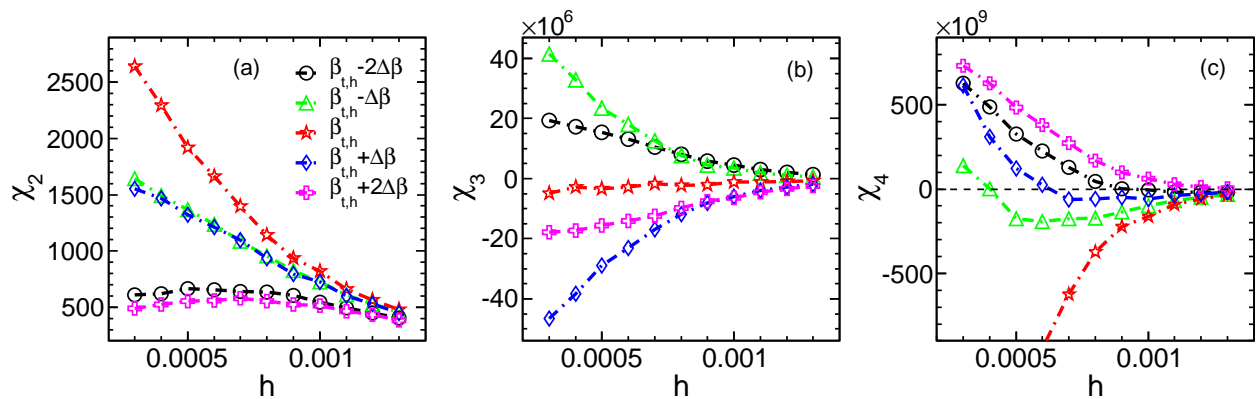}
	\caption{\label{Fig. 3}(color online) $\chi_2$(a), $\chi_3$(b) and $\chi_4$(c) on the lines above, along and below the phase boundary ($\beta_{t,h}$).}
\end{figure*}

In contrast, as shown in Fig. 2(b) and 2(d), the sign of $\chi_4$ changes twice, and that of $\chi_6$ four times. The phase boundary is located in a yellow for $\chi_4$ and a green band $\chi_6$. Far away from the phase boundary, their signs both remain positive in the two phases.

So, over the whole phase space, the sign change happens only in the vicinity of the phase transition line. Far away from the phase transition line, the sign of odd-order susceptibility remains positive and negative in the symmetric and broken phases, respectively, while that of even-order susceptibility is  positive in both phases. The sign of odd-order susceptibility at high (low) temperature is associated with symmetry (broken) phase in the Potts model.

The effective area of the phase transition related fluctuations is relevant to the external field, and the system size. The sign distributions of $\chi_4$, $\chi_5$, and $\chi_6$ in Fig. 2(b), (c) and (d) show that the width of the bands changes with the external field. It becomes wider and wider with the increase of the external magnetic field. The bigger the external field, the wider the width of the bands.

On the other hand, the finite system size makes the fluctuation remain in a small temperature region in the vicinity of the phase transition. The larger the system size, the narrower the width of the bands, cf., Fig.~2 to Fig.~6 of Ref.~\cite{Xue-JPG42}. For an infinite system, the fluctuation is divergent, and the width of the fluctuation area is zero.

In order to analyze the behavior of the generalized magnetic susceptibilities clearly when the critical point is approached from the crossover side, four lines, as well as the phase boundary itself, are chosen near the phase boundary in the $\beta-h$ plane. They are $\beta_{t,h}\pm\Delta\beta$, with $\Delta\beta=0.00005$, and $\beta_{t,h}\pm2\Delta\beta$, where $\beta_{t,h}=1/T_{t,h}$ denotes the reciprocal of the phase transition temperature at the corresponding external magnetic field. On each line, we have chosen eleven values of $h$ and calculated the corresponding second-, third- and fourth-order magnetic susceptibilities, respectively. The results are shown in Fig.~3.

These five lines are presented, from top to bottom, by black circles, green triangles, red stars, blue diamonds and violet crosses in Fig.~3(a), 3(b) and 3(c), for $\chi_2$, $\chi_3$ and $\chi_4$ respectively. $\chi_2$ on the phase boundary (red stars) decreases monotonically with the external field, as do the green triangles and blue diamonds with temperature deviating $\Delta\beta$ from the phase transition. When the temperature deviates by $2\Delta\beta$ from the phase transition temperature, i.e., the black circles and violet crosses, there seems to be a small peak. So even if on the phase boundary, or very close to it, the non-monotonic fluctuations are hard to observe in the second-order magnetic susceptibility.

However, $\chi_3$ in Fig.~3(b) on the phase boundary stays around zero, the same as shown in Fig.~2(a). When the temperature deviates by $2\Delta\beta$ and $\Delta\beta$ ($-2\Delta\beta$ and $-\Delta\beta$) from the phase transition temperature,  $\chi_3$ increases (decreases) monotonically with external field. So,
along the lines near the phase boundary, sign change and non-monotonic fluctuations are not observable in $\chi_3$.

Along the phase boundary, $\chi_4$ is negative and increases monotonically with external magnetic fields, as shown by the red stars in Fig.~3(c). When the temperature deviates  by $\Delta\beta$ from the phase transition temperature, $\chi_4$ is negative when the critical point is approached from the crossover side, which is consistent with the prediction in Ref.~\cite{Stephanov-PRL107}. The sign of $\chi_4$ changes at the first-order phase transition region, as shown by the green triangles and blue diamonds. When it deviates by $2\Delta\beta$ from the phase transition temperature, as shown by the black circles and violet crosses in Fig.~3(c), the values of $\chi_4$ are mostly positive in the  interval of $h$ studied. Therefore, when it is not far from the phase transition line, the sign change and non-monotonic fluctuations are still observable in $\chi_4$.

\section{Finite-size behavior of generalized susceptibilities of energy}

In systems of sizes $L=40$, 50, 60 and 70, the temperature dependence of the second- to fourth-order susceptibilities of the energy is studied at the critical external magnetic field $h=h_c$. The results are shown in Fig.~4. It is clear that $\chi_2^E$ has a peak for each size in Fig.~4(a). As the size increases, the peak becomes higher and its position shifts to the higher temperature side. $T_c$ is set by the temperature of the peak with system size $L=70$ and its reciprocal $1/T_c=0.549395$.

\begin{figure*}
	\includegraphics[width=0.95\textwidth]{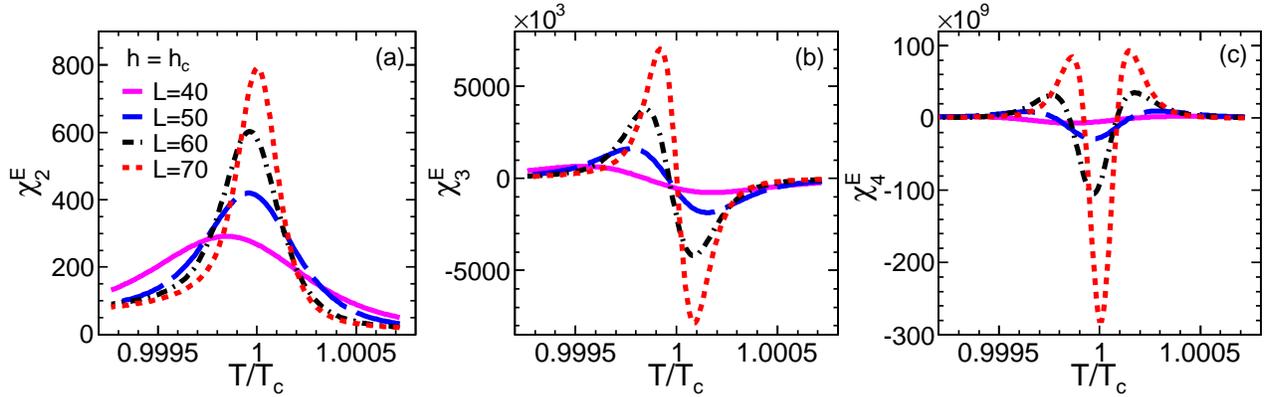}
		\caption{\label{Fig. 4}(color online) The temperature dependence of $\chi_2^E$(a), $\chi_3^E$(b) and $\chi_4^E$ (c) at the critical external magnetic field $h=h_c$. }
\end{figure*}

The temperature dependence of $\chi_3^{E}$ has an oscillation structure. In the vicinity of $T_c$, its sign changes from positive to negative with increasing temperature. For $\chi_4^{E}$ in Fig.~4(c), it oscillates more violently. Its sign changes twice. Near $T_c$, there is a deep negative valley. In different sizes of systems, the qualitative structure of $\chi_3^{E}$ or $\chi_4^{E}$ is the same.

The finite-size scaling law for second-order phase transitions is well established in statistical physics~\cite{second scaling, Chen XS-PA232}. In the vicinity of the critical point, the specific heat, $c=\chi_2^E/T^2$, is a function of the temperature and system size. It follows the finite-size scaling relation~\cite{Privman-1990}
\begin{equation}\label{scaling-chi2}
c=L^{\alpha/\nu}F(t L^{1/\nu}),
\end{equation}
where $t=\frac{T-T_c}{T_c}$ is the reduced temperature. $\alpha$ and $\nu$ are the critical exponents of the specific heat and the correlation length, respectively.
$F(t L^{1/\nu})$ is the scaling function of scaled variable $t L^{1/\nu}$. At the critical point $t=0$, the scaling function $F(t L^{1/\nu})$ is a constant independent of $L$. So
\begin{equation}\label{scaling relation}
\chi_2^E(L) \propto L^{\alpha/\nu}.
\end{equation}

If we take the logarithm of Eq.~\eqref{scaling relation}, then
\begin{equation}\label{logarithm-chi2}
\ln \chi_2^E(L)= (\alpha/\nu)\ln L+C_1,
\end{equation}
where $C_1$ is a constant. At the critical point, the double logarithm plot of $\chi_2^E$ versus $L$ is a straight line. When the temperature deviates from the phase transition, the scaling function is system size dependent. $\ln\chi_2^E$ is no longer a linear function of $\ln L$.

At $h=h_c$ and $\beta=1/T_c=0.549395$, the log-log plot of $\chi_2^E$ versus $L$ is shown by the red line in Fig.~5. It is a straight line, as expected. When the temperature deviates from $T_c$, the log-log plot is not a straight line, just as the blue and black lines show.

The slope of the red line is 1.93164 $\pm$ 0.02.
This is quite different from the critical exponent ratio $\alpha/\nu \approx0.1746$ in the three-dimensional Ising universality class, but approximates the ratio $\lambda/\nu\approx1.963$, where $\lambda$ is the critical exponent of the magnetic susceptibility~\cite{scaling exponent}.

$\chi_2^E$ is the second-order derivative of the free energy to $\beta$. From Eq.~\eqref{linear relation}, $\beta$ is related to both $\tau$ and $\zeta$. The critical exponent $y_{\zeta}\approx2.5$ is bigger than $y_{\tau}\approx 1.6$~\cite{scaling exponent}, so the finite-size scaling of $\chi_2^E$ is mainly controlled by the derivatives of the free energy to $\zeta$, which is why the slope of the red line in Fig.~5 approximates the ratio $\lambda/\nu$ but not $\alpha/\nu$.

\begin{figure}
	\includegraphics[width=0.42\textwidth]{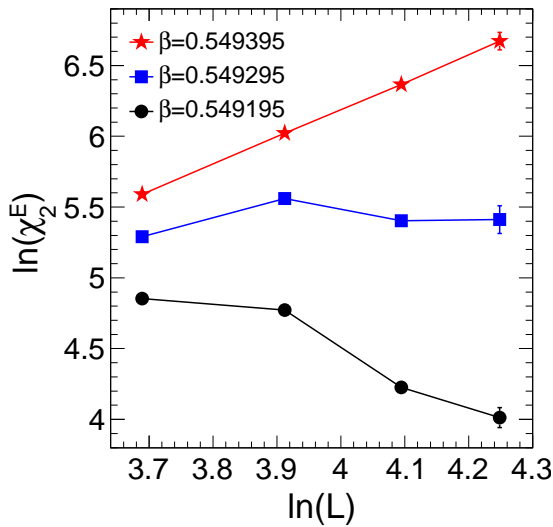}
		\caption{\label{Fig. 5}(color online) The log-log plot of $\chi_2^E$ and $L$ at the critical external magnetic field and three different values of $\beta$.}
\end{figure}

\section{Summary and conclusions}

Using the three-dimensional three-state Potts model, we have studied the generalized magnetic susceptibilities. Their sign distribution in the temperature-external magnetic field plane has been presented. The behavior of the magnetic susceptibilities has been discussed when the critical point is approached from the crossover side. The temperature dependence of the second- to fourth-order susceptibilities of the energy in different sizes of systems has also been shown and discussed.

For the generalized magnetic susceptibilities, their sign change happens only in the vicinity of the phase boundary. The effective area of the phase transition related fluctuations is dependent on the external field and system size. The bigger the external field and the smaller the system size, the wider the area is.

Far away from the phase boundary, the sign of odd-order susceptibility is positive and negative in the symmetric (disorder) and broken (order) phases, respectively, while the sign of the even-order susceptibility remains positive in both phases.

When the critical point is approached from the crossover side, the second- and third-order of susceptibilities remains monotonic, while the negative values of the fourth-order susceptibility are still observable in the vicinity of the critical point.

At the critical external magnetic field, the peak structure, oscillation or sign change can be observed in the temperature dependence of the susceptibilities of the energy. Only at the critical temperature, the double logarithm plot of the second-order susceptibility of the energy versus the system size is a straight line. If the temperature is away from the critical temperature, the plot deviates from a straight line. From the slope of the straight line, the finite-size scaling behavior of the specific heat in the three-dimensional three-state Potts model is mainly controlled by the critical exponents of the magnetic susceptibility in the three-dimensional Ising universality class.

This work is supported by the NSFC of China under Grants No. 11647093, 11405088 and 11521064, Fund Project of Chengdu Technological University under Grant No. 2016RC004, the Major State Basic Research Development Program of China under Grant No. 2014CB845402.

\ed